\documentstyle[amssymb]{mn}

\title[A Virgo-Fornax distance modulus]{Some effects of galaxy structure and dynamics on the Fundamental Plane.\ II.\ A Virgo-Fornax distance modulus}
\author[A.W.Graham]{Alister W.\ Graham\thanks{E-mail: ali@mso.anu.edu.au} \\
	Mount Stromlo and Siding Spring Observatories, The Australian National University, \\Private Bag, Weston Creek PO, ACT 2611, Australia.}
\date{ }

\begin{document}

\input{psfig}
\psrotatefirst

\maketitle

\begin{abstract}

The influence of broken structural homology, and the implied broken dynamical 
homology, is examined for the Fundamental Plane (FP). 
Requiring a symmetrical treatment of the FP variables, 
a bisector method of linear regression was applied, in 3-dimensions, 
to derive the best FP.  A bootstrapping procedure
has been used to estimate the uncertainties associated with the slope of the FP. 
For 25 E and S0 Virgo galaxies, 
the `standard' FP, constructed using $R^{1/4}$ model parameters for the 
effective radii ($R_{\rm e,4}$) and the mean surface brightness within this 
radius ($\Sigma _{\rm e,4}$) and using central velocity dispersion 
(CVD) measurements ($\sigma _0$), gave a relation described by 
$R_{\rm e,4}\propto \sigma _{0}^{1.10\pm0.14}\Sigma _{\rm e,4}^{-0.55\pm0.09}$.
Using Sersic $R^{1/n}$ light profile model parameters and the projected, infinite 
aperture, velocity
dispersion ($\sigma _{\rm tot,n}$), derived from application of the 
Jeans equation to the observed intensity profiles, gave an `improved' FP 
described by the relation 
$R_{\rm e,n}\propto \sigma _{\rm tot,n}^{1.37\pm0.16}\Sigma _{\rm e,n}^{-0.76\pm0.05}$.  
This result, based on independent data, supports the previous finding
by Graham \& Colless (1997a) that assumptions of 
structural and dynamical
homology are partly responsible for the departure of the
observed FP from the plane expected by the virial theorem, which predicts 
$R\propto \sigma ^{2}\Sigma ^{-1}$.   
Upon removal of the known S0 galaxies from the sample of Virgo galaxies, 
the above planes were observed to change to 
$R_{\rm e,4}\propto \sigma _{0}^{1.19\pm0.21}\Sigma _{\rm e,4}^{-0.60\pm0.11}$
and 
$R_{\rm e,n}\propto \sigma _{\rm tot,n}^{1.72\pm0.24}\Sigma _{\rm e,n}^{-0.74\pm0.09}$.  
The perpendicular rms residuals about these planes are 0.084 and 0.050 dex, respectively.  

The Fornax cluster was similarly treated, although removal of the S0 
galaxies left a sample of only 7 ellipticals which had published CVD 
measurements.  Treating the range of structural and dynamical profiles 
present in this sample produced a FP given by the relation 
$R_{\rm e,n}\propto \sigma _{\rm tot,n}^{2.03\pm0.78}\Sigma _{\rm e,n}^{-1.07\pm0.30}$,
in tantalizing agreement with the plane expected from the virial theorem, but
with discouragingly large errors due to the small sample size.
Similarly to Virgo, the perpendicular rms residuals about this plane is 
0.050 dex.  

The FP was also constructed with the purpose of using it as a distance indicator, 
achieved by minimising the distance-dependent quantity $\log R$ against 
the distance-independent quantities $\Sigma $ and $\log \sigma$.
A Virgo-Fornax distance modulus was computed using 
Working-Hotelling confidence bands (Feigelson \& Babu 1992).  
The `standard' FP parameters gave a value of 0.45$\pm$0.16 mag, 
where-as the `improved' FP parameters gave a value of 
0.25$\pm$0.12 mag. 
However, 
a full treatment of the uncertainties on the FP slopes, derived through 
a bootstrapping procedure of the 3-dimensional FP data set, revealed that 
the analytical expressions for the uncertainties on the estimated distance 
moduli, given above, should be increased by a factor of $\sim$5.

\end{abstract}


\section{Introduction}

Our knowledge about the structure of elliptical galaxies has 
progressed considerably in the last decade.
No longer are we satisfied with fitting a de Vaucouleurs (1948,1953) 
$R^{1/4}$ intensity distribution
to the light profiles of elliptical galaxies.  Modern imaging techniques are 
providing ever-more precise measurements, revealing a range of light profile 
shapes which have been shown to be better described by the Sersic (1968) 
$R^{1/n}$ 
model (Capaccioli 1989; Caon, Capaccioli \& D'Onofrio 1993; Graham et al.\ 
1996, and references within).  
Through varying the parameter $n$ in the Sersic model, 
one can reproduce an exponential light distribution ($n$=1), the classic
de Vaucouleurs profile ($n$=4), approximate a power-law for large values 
of $n$ (approximately $>$15), 
or any of the intermediate forms 
which galaxies exhibit (Michard 1985; Schombert 1986; Binggeli \&
Cameron 1991; Caon et al.\ 1993; Graham \& Colless 1997a, hereafter 
Paper I, to mention a few).  
Consequently, for those galaxies which are not well described by an 
$R^{1/4}$ profile, the effective radius and mean surface brightness within 
this radius will be different when obtained using the de Vaucouleurs model
and the Sersic model. More importantly, these differences 
have been shown to have a systematic trend: 
galaxies with $n$$<$4 have their effective half-light radii over-estimated 
by the $R^{1/4}$ model, and galaxies better described with values of $n$$>$4 
have their effective half-light radii under-estimated by the $R^{1/4}$ model. 
Similarly, the mean surface brightness is under- and over-estimated by the 
$R^{1/4}$ model when $n$$<$4 and $n$$>$4 respectively (Graham \& Colless
1997b).

Observed variations in the galaxy structure, as described by the $R^{1/n}$ 
model, imply a variety of dynamical structures (Ciotti 1991; Paper I) 
are required to 
maintain these predominantly pressure-supported stellar systems.  
This conjecture is borne out in the observations of elliptical galaxy
dynamics (Davies et al.\ 1983; Illingworth 1983; Capaccioli \& Longo 1994) 
and has implications for the construction of the Fundamental Plane (FP). 
In particular, the use of a galaxy's central velocity dispersion 
(CVD, $\sigma _0$) to represent the overall kinetic energy of a galaxy has 
been shown to be inappropriate, 
(J\o rgensen, Franx \& Kj\ae rgaard 1993; Bender, Saglia \& Gerhard 1994).
The velocity dispersion derived from the 
kinetic energy of random motion over the entire galaxy surface has been 
shown to be proportional to 
$\sigma _0^{1.6}$ rather than $\sigma _0^{2}$ (Busarello et al.\ 1997). 
A procedure is described in Paper I, based upon the theoretical study 
of Ciotti (1991), which enables the velocity dispersion to be measured 
in a consistent fashion between galaxies.  
Rather than simply using CVD measurements, that sample different fractions
of different galaxies' velocity dispersion profiles (in terms of the half-light radii),
through application of the Jeans equation, Paper I presented a method to 
compute the asymptotic value of the velocity dispersion within a projected
aperture of infinite radius.  This measure, denoted by $\sigma _{tot}$, 
has the added significance of being equal to one-third 
of the virial velocity for spherical systems (Ciotti 1994).

In Paper I the light profile data (aperture 
magnitudes from Bower, Lucey \& Ellis 1992) from a sample 
of 26 E/S0 Virgo galaxies was analysed.  
Departures from the $R^{1/4}$ model were found to be well-described
by the Sersic model, and the implied non-homology in the velocity
dispersion profile was explored.  Using the Sersic model parameters,
rather than $R^{1/4}$ model parameters, and using the computed, 
infinite aperture velocity dispersion rather 
than the CVD, the FP was observed to change from  
$R_{\rm e,4}$$\propto$$\sigma _{0}^{1.33\pm0.10}\Sigma _{\rm e,4}^{-0.79\pm0.11}$
to 
$R_{\rm e,n}$$\propto$$\sigma _{\rm tot,n}^{1.44\pm0.11}\Sigma _{\rm e,n}^{-0.93\pm0.08}$.  
Broken structural and dynamical homology appeared to be responsible 
for some of the {\it tilt} of the FP, i.e.\ its departure from the plane 
predicted by the virial theorem, $R$$\propto$$\sigma ^{2}\Sigma ^{-1}$.

Ultimately, this broken dynamical and structural homology needs to be 
dealt with if one is to understand the nature of elliptical galaxies
and use them for measuring distances (Dressler et al.\ 
1987). 
Given the substantial improvements in describing elliptical galaxies, 
it is apt that the FP, and its application to distance estimates, is
revisited.  
Using a new data base (taken from Caon et al.\ 1994) obtained by different 
observers with a different instrument and using a different reduction 
procedure, the influence of broken structural and broken dynamical homology
upon the FP is again investigated.  The data used
in this paper is the surface brightness profiles along the major axis of each 
galaxy, as opposed to the integrated aperture magnitudes used in 
Paper I.   
The influence of S0 galaxies upon the slope of the FP
is also investigated, with comparison of the FPs constructed with, and without, 
their inclusion. 

This paper presents the FP for the galaxy clusters
Fornax and Virgo, two key clusters in the extra-galactic distance scale
(Jacoby et al.\ 1992; McMillan, Ciardullo \& Jacoby 1993; Bureau, 
Mould \& Staveley-Smith 1996). 
These clusters are sufficiently distant to possess significant 
motion due to the Hubble expansion and yet they are close enough that their
distances may be obtained with a plethora of independent observational 
techniques.  
The following section presents the galaxy sample and the observational/model 
data which is used in this study. The Fundamental Plane is computed, and 
its method of
construction, including a new error analysis routine, is described 
in Section 3.  Using two different methods of analysis, a distance 
modulus between Virgo and Fornax is calculated in Section 4. 
This result is compared with other estimates of the Virgo-Fornax distance 
modulus in Section 5, along with a discussion of the other results.  
Section 6 provides a summary and concluding remarks.

\section{Galaxy Data}
\subsection{The galaxy sample}

The galaxy sample from Caon et al.\ (1994) has been used for this 
investigation. 
This data set has been used in a benchmark study (Caon et al.\ 1993) 
which gave insight into the 
nature of elliptical galaxy light profiles by quantifying their 
deviations from the $R^{1/4}$ model.  Knowing 
that this galaxy sample does not posses structural homology amongst its
members makes it an ideal sample to explore the
necessarily implied broken dynamical homology and the combined influence 
of these effects upon the FP.  From the sub-sample of galaxies for which 
major-axis profiles have been modeled 
(see Caon et al.\ 1993 for a discussion of some of the problems in 
doing this), those which have no published velocity dispersion data 
have been excluded here. 
This reduced the sample of Virgo galaxies from 33 (Caon et al.\ 1993) 
to 25, and the sample of Fornax galaxies from 16 (D'Onofrio, Capaccioli 
\& Caon 1994) 
to 10.   The 10 Fornax galaxies were further reduced to a sample of 9 
after the exclusion of NGC 1399, which, with its shallow light profile
($n$$>$15), could not have its velocity dispersion profile accurately modeled
due to the large uncertainties associated with the effective radius $R_e$.  
The remaining galaxies are listed in Table 1.
%

\subsection{Surface brightness profiles and structural model parameters}

The B band surface brightness profiles from Caon et al.\ (1994) have been 
used in this study.  They were constructed by these authors using the 
`global mapping' procedure of Capaccioli \& Caon (1989) which
couples CCD data for the bright inner galaxy profile with photographic
data for the outer profile.  This technique allows very deep
photometry of galaxies, and alleviates the problem of uncertain 
sky subtraction errors that is associated with the use of small area CCD's.
The method gives internal errors of less than 0.1 mag at $\mu _B$=26 mag.  

In this study the major-axis surface brightness profiles have been
truncated at the same inner and outer limits that were applied by Caon et 
al.\ (1993) and D'Onofrio et al.\ (1994) for the Virgo and Fornax 
galaxies respectively.  In general, this meant fitting the entire light 
profile after the exclusion of the inner portion affected by seeing, and 
the outer portion where sky subtraction uncertainties became 
large ($>$0.25 mag). This left a typical range in surface brightness from 
$\mu _B$=20 to 27 mag.

The Sersic (1968) $R^{1/n}$ light profile model was fitted to the remaining
profile, with the intensity $I$ given as a function of radius $R$ such that 
\begin{equation}
I(R)=I_{e}exp\left[ -b(n)\left[\left( \frac{R}{R_{e}}\right) ^{1/n} -1\right]\right].
\end{equation}
$I_{e}$ is the intensity at the radius $R_{e}$, and 
$b(n)$ ($\approx$1.9992n-0.3271, Capaccioli 1989)
is defined such that $R_{e}$ is the radius that effectively encloses half 
of the total light of the profile model (Caon et al.\ 1993; 
Graham \& Colless 1996).  The `shape parameter' $n$ describes the curvature
of each light profile.  Application of a non-linear least-squares fit of 
the $R^{1/n}$ model to the galaxy 
surface brightness data, with equal weight for each data point, produced 
similar values for the shape parameters to those derived by Caon et al.\ 
(1993) and D'Onofrio et al.\ (1994), confirming that the models are being 
fitted consistently between authors. 

Table 1 lists the parameters from the best fitting $R^{1/n}$ 
model and also those from the best fitting $R^{1/4}$ model, which was fitted 
in an identical fashion.  
$\Sigma _e$ is the mean surface brightness within the 
effective radius $R_e$, computed using the technique described
in Appendix A of Paper I.  
The subscripts $n$ and 4 on the model parameters 
are used to denote the model from which the particular parameters have come; 
$n$ signifying $R^{1/n}$ model parameters, and 4 signifying $R^{1/4}$ model
parameters.   

Ciotti (1991), and also Paper I, explored the relationship between the 
projected and spatial (deprojected) quantities.  By definition, the surface 
brightness terms had a one-to-one mapping, while there was an approximately 
constant offset ($\log R$$\sim$0.13, Ciotti 1991) between the 
effective radii.  Therefore either of these quantities could be used for the
construction of the FP.
The deprojected velocity dispersion at
the effective radius was found not to have such a simple relation
with its projected counterpart, the central velocity dispersion measurement.
This deprojected velocity dispersion is preferred over the CVD measurement
because it not only allows
for the range of velocity dispersion profiles present, but
also samples different galaxies in a consistent fashion
(CVD measurements sample typically anywhere between 0.01-0.2 $R_e$).
However, due to the physical significance of the projected velocity
dispersion within an aperture of infinite radius, this term, which also
samples different galaxies in a consistent fashion, is preferred
in the analysis which follows.

\begin{figure}
\centerline{\psfig{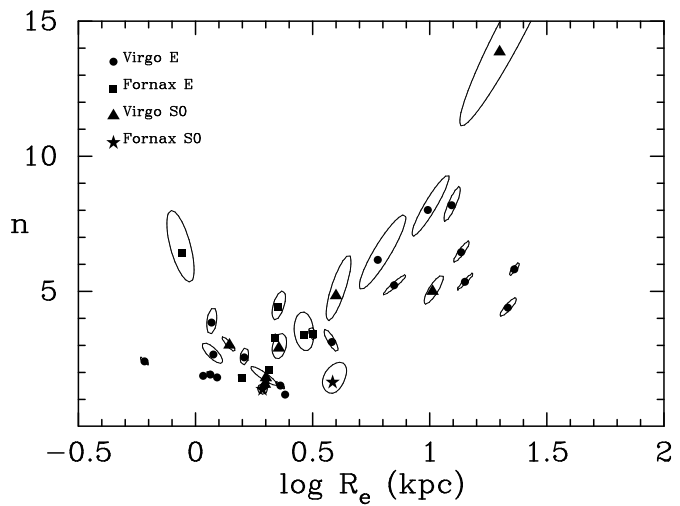}}
\caption{The shape parameter $n$ is plotted against the $R^{1/n}$
model effective radius $R_e$ for the Virgo and Fornax galaxies listed in
Table~\ref{tab1}. 
The associated 99\% confidence interval is also drawn, as detailed in 
the text.}
\label{fig1}
\end{figure}

Following Caon et al.\ (1993) (their Fig.\ 3), the light profile shape 
parameter is plotted against galaxy scale-size in Figure~\ref{fig1}. 
Although our methods of parameterisation differ (Caon et al.\ 1993 have 
used model-independent estimates for the half-light radius), the same trend 
of increasing values of $n$ with $R_e$ is evident here.
The present study also includes Fornax galaxies as well as those from Virgo. 
Also shown in the figure are the 99\% confidence intervals for the model parameters.
These were computed by evaluating the chi-square values for a fine grid of 
points around the optimum solution.  The contours drawn correspond to 
$\Delta \chi ^{2}$=9.21 (i.e.\ 4$\sigma $), scaled after 
normalisation of the reduced $\chi ^{2}$ of the optimum solution. 
\footnote{It is noted that these confidence intervals only encompass the 
uncertainty of the parameters derived from the analytical fit to the 
data.  They do not incorporate measurement errors or zero point offsets 
which may be present in the data.}

\begin{figure}
\centerline{\psfig{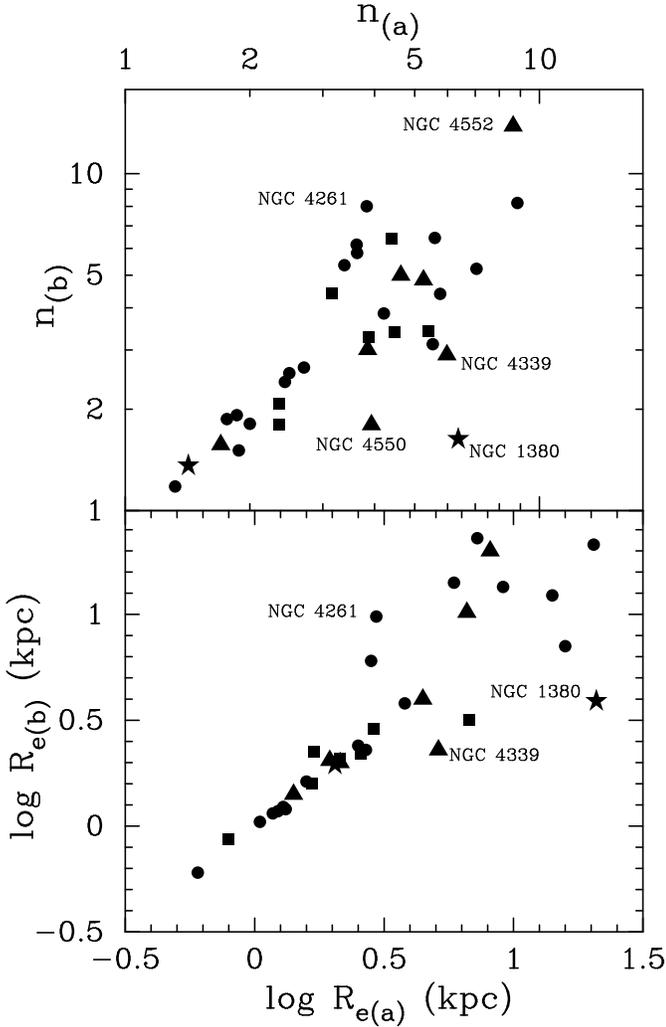}}
\caption{The Sersic $R^{1/n}$ model parameters from the fit to the entire 
light profile (subscript b) are plotted against those parameters obtained
from fit to the light profile within 40$\arcsec$ (subscript a).  
The circles represent Virgo E's (18), the triangles are Virgo S0's (7), 
the squares are Fornax E's (7) and the stars represent Fornax S0's (2).}
\label{fig2}
\end{figure}

In exploring the sensitivity of the $R^{1/n}$ model parameters to the 
radial range of the surface brightness profile, the Sersic 
model was fitted to that data within the inner 40$\arcsec$. 
This outer truncation is the same as the outermost data points used in 
Paper I, enabling a further comparison of the present data with 
that used in Paper I.  The inner truncation  used by Caon et al.\ 
(1993) and D'Onofrio et al.\ (1994) was again employed, except for NGC 
4473 and NGC 4636 for which only the inner 3$\arcsec$ was excluded. 
In Figure~\ref{fig2}, the resulting model parameters are shown plotted 
against those obtained using the entire surface brightness profile (Table 1). 
It is noted that the light profiles from the smaller radial extension 
($<$40$\arcsec$) may not be representative of the galaxy as a whole, and 
therefore may not represent the {\it over-all} global galaxy shape.  This is
more likely to be true for the larger galaxies, especially those
having $R_e$$>$40$\arcsec$.  It is also true that the Sersic index $n$ 
is less sensitive at larger values (Graham et al.\ 1996).  Thus, 
Figure~\ref{fig2} offers an alternative estimate for the uncertainty 
of the $R^{1/n}$ model parameters.  Some of the more discrepant points have 
been labeled in the plot.  It is noted that they are predominantly 
the S0 galaxies.  If one is to think of these galaxies as possessing
an $R^{1/4}$ ($n$=4) central bulge and an exponential ($n$=1) 
outer disk, or some approximately equivalent Sersic form, 
then this could account for the larger index obtained using the smaller 
radial range for the S0 galaxies NGC 1380, 4339, and 4550. 
In fact, such measurements offer themselves as possible diagnostics for 
distinguishing S0 galaxies from ellipticals.

For the Fornax galaxies, the tabulated extinctions range from $A_B$=-0.05
to -0.09.
Some of the Virgo galaxies also have negative extinctions, with a
mean Virgo cluster value of $A_B$=0.04$\pm$0.07 (standard deviation) for the
sample.  The negative values are due to the variable ratio of H$_{I}$
gas-to-dust in the Galaxy (Burstein \& Heiles 1978), and have no
physical meaning. 
Such values arise in the method used by Burstein \&
Heiles for regions where there is zero or negligible reddening,
and they should be set equal to 0.00.
Corrections for Galactic extinction (Burstein \& Heiles 1984; Burstein
et al.\ 1987) were not applied by Caon et al.\ (1994).
This practice has been maintained here.

At a redshift of around 0.003, the k-correction and redshift dimming
will be equally small for both clusters, at 0.015 mag and 0.013 respectively.  
Consequently these terms have not been incorporated into the analysis.

\subsection{Dynamical model parameters}

CVDs are typically used as a measure of the kinetic 
energy within elliptical and S0 galaxies.
There are many problems associated with doing this.  Firstly, rotational
energies, which are likely to be significant for S0 galaxies, 
are ignored.  Secondly, fixed aperture sizes (on the sky) 
in which the CVD is measured correspond to different physical sizes 
for galaxies in clusters at different distances.  This effect alone
may likely introduce systematic errors into cluster distance determinations
which have incorporated fixed aperture sizes to measure the CVD.  
Thirdly, given that
galaxies have a range of sizes and structural profiles, they will also have 
a range of dynamical profiles which support the various structures that are 
observed.  Therefore, even within clusters, fixed aperture sizes will 
sample different fractions of individual galaxies, as measured 
in terms of their effective radii.  

In an attempt to find a better measure of the kinetic energy within each
galaxy, a method of computing the velocity 
dispersion in a consistent manner for all galaxies was presented in Paper I. 
Deprojecting the galaxy light profile, the spatial luminosity density (Ciotti 
1991) was computed.  Assuming a constant $M/L$ ratio within each galaxy, 
such knowledge of the internal density profile coupled with the application of
the Jeans hydrodynamical equation enables one to compute the internal
velocity dispersion profile.  This can then be projected outward to
give the projected radial velocity dispersion profile, which in turn can
be integrated over increasing aperture sizes to give 
the projected aperture velocity dispersion profile.  This is then calibrated
using the measured CVD.

The computed spatial (deprojected) velocity dispersion at the spatial 
half-light radius was used in Paper I to represent 
the kinetic energy of the galaxies.  Unlike use of CVD, this 
approach consistently samples the velocity dispersion profile in the 
same way for all galaxies.  As an alternative, the computed velocity 
dispersion within an aperture of infinite radius was also used.  This 
quantity has been shown by Ciotti (1994) to be equal to one-third
of the virial velocity dispersion, independent of any orbital 
anisotropies that may be present.  This measure for the kinetic energy is 
adopted here, not only for the consistent fashion in which galaxies are 
sampled but also due to the significance of the infinite aperture velocity 
dispersion $\sigma _{tot}$.  These values are listed in Table 1, 
where the subscripts 4 and $n$ refer to use of the $R^{1/4}$ and $R^{1/n}$
models respectively.  Also listed are the CVD measurements, $\sigma _0$, 
from McElroy (1995).

\begin{table*}
\begin{minipage}{110mm}
\caption{Galaxy model parameters.  The galaxy sample is a selection from 
Caon et al.\ (1994).  
The parameters are from the best fitting $R^{1/4}$ and $R^{1/n}$ models,
with the exception of $\sigma _{0}$ which is the tabulated central velocity
dispersion from McElroy (1995).
$R_{e,4}$ and $R_{e,n}$ are the effective radii of the $R^{1/4}$ 
and the $R^{1/n}$ models respectively, with $\Sigma _{e,4}$ and 
$\Sigma _{e,n}$ the mean surface brightness within $R_{e}$.  
$\sigma _{\rm tot,4}$ and $\sigma _{\rm tot,n}$ are the 
infinite aperture velocity dispersions described in the text.}
\label{tab1}
\begin{tabular}{llrrrcrrcr}
\hline
Galaxy&Morph.&$\sigma _{0}$&\multicolumn{1}{c}n&$\log R_{\rm e,n}$&$\Sigma _{\rm e,n}$&$\sigma _{\rm tot,n}$&$\log R_{\rm e,4}$&$\Sigma _{\rm e,4}$&$\sigma _{\rm tot,4}$\\[.2ex]
Ident.&Type&km s$^{-1}$& &[kpc]&[mag]&km s$^{-1}$&[kpc]&[mag]&km s$^{-1}$\\[10pt]
NGC 1339 &E4    &160 &1.80 &0.20 &20.78 &138 &0.04 &19.74 &134\\ 	
NGC 1351 &E5    &143 &3.38 &0.46 &21.52 &112 &0.46 &21.46 &110\\
NGC 1374 &E0    &186 &3.28 &0.34 &20.93 &149 &0.32 &20.77 &147\\
NGC 1375 &S0    & 53 &1.36 &0.29 &21.27 & 53 &0.40 &21.59 & 41\\
NGC 1379 &E0    &119 &2.07 &0.32 &20.96 &100 &0.26 &20.57 & 95\\
NGC 1380 &S0/a  &225 &1.63 &0.59 &20.61 &191 &0.72 &21.01 &166\\
NGC 1404 &E2    &250 &4.42 &0.35 &19.82 &193 &0.34 &19.79 &196\\
NGC 1419 &E0    &133 &6.41 &-0.06&20.35 &110 &0.02 &20.89 &112\\
NGC 1427 &E4    &155 &3.42 &0.50 &21.36 &121 &0.49 &21.27 &118\\
NGC 4168 &E2    &186 &6.16 &0.78 &22.82 &125 &0.64 &22.35 &139\\
NGC 4261 &E2    &326 &8.01 &0.99 &22.82 &196 &0.74 &21.91 &240\\
NGC 4339 &S0(0) &114 &2.90 &0.36 &21.48 & 92 &0.38 &21.54 & 89\\
NGC 4365 &E3    &261 &6.45 &1.13 &22.83 &164 &0.96 &22.14 &190\\ 
NGC 4374 &E1    &296 &8.19 &1.09 &22.33 &173 &0.91 &21.69 &216\\
NGC 4387 &E5    &112 &1.92 &0.06 &20.35 & 99 &0.00 &19.89 & 95\\
NGC 4431 &dS0,N & 68 &1.57 &0.30 &22.29 & 58 &0.60 &23.50 & 51\\
NGC 4434 &E0    &115 &3.85 &0.07 &20.63 & 96 &0.07 &20.63 & 96\\
NGC 4458 &E1    &106 &2.56 &0.21 &21.30 & 89 &0.24 &21.29 & 85\\
NGC 4459 &S0(2) &178 &4.84 &0.60 &21.20 &130 &0.57 &21.14 &134\\
NGC 4464 &E3    &121 &2.41 &-0.22&19.48 &111 &-0.38&18.44 &112\\
NGC 4472 &E2    &303 &5.82 &1.36 &22.74 &193 &1.21 &22.15 &221\\
NGC 4473 &E5    &193 &3.12 &0.58 &20.88 &151 &0.53 &20.51 &146\\
NGC 4476 &S0(5) & 81 &3.01 &0.15 &20.77 & 68 &0.06 &20.19 & 68\\
NGC 4478 &E2    &143 &1.87 &0.03 &19.70 &127 &-0.10&18.92 &124\\
NGC 4486 &E0    &333 &5.35 &1.15 &22.29 &221 &1.01 &21.78 &242\\
NGC 4550 &S0(7) & 83 &1.80 &0.30 &20.25 & 71 &0.01 &18.38 & 70\\
NGC 4551 &E2    &113 &1.81 &0.09 &20.56 & 99 &-0.03&19.90 & 96\\
NGC 4552 &S0(0)\footnote{Also known as M 89, this galaxy is catalogued as type E in de Vaucouleurs et al.\ (1991).} &269 &13.86&1.30 &23.73 &117 &0.76 &21.60 &198\\
NGC 4564 &E6    &160 &1.51 &0.36 &20.59 &137 &0.07 &18.63 &133\\
NGC 4621 &E4    &230 &5.22 &0.85 &21.61 &159 &0.70 &21.05 &170\\
NGC 4623 &E7    & 89 &1.18 &0.38 &21.41 & 89 &0.30 &20.71 & 70\\
NGC 4636 &E1    &207 &4.40 &1.33 &23.43 &147 &1.30 &23.29 &152\\
NGC 4649 &S0(2) &339 &5.00 &1.01 &21.70 &232 &0.95 &21.48 &247\\
NGC 4660 &E3    &185 &2.66 &0.08 &19.39 &159 &-0.03&18.78 &158\\
\hline
\end{tabular}
\end{minipage}
\end{table*}

\section{Construction of the Fundamental Plane}
\label{sectc}

The quantities which are used in this investigation of the FP are
the effective radius, $R_e$, and the mean surface brightness within this
radius, $\Sigma _e$, derived from the model fits to the surface brightness
profiles.
These values are listed in Table 1. 
This approach differs from D'Onofrio, Longo \& Capaccioli (1996), who used 
model-independent values for these quantities in their construction of the 
FP.  They also only used CVDs for their velocity dispersion measure.

In this section, the Fundamental Plane is 
constructed with the goal of comparison with theoretical predictions. 
Consequently, the 3 parameters which are used to define the FP 
($\sigma $,$R$,$\Sigma $) are treated as independent of each other.
This requires a symmetrical treatment of these parameters, and 
the bisector method of linear regression is the preferred technique 
(Isobe et al.\ 1990; Feigelson \& Babu 1992). This approach 
requires that three FPs be computed, each of which is 
obtained by minimising the residuals of a different parameter.  
The mean slope angles of these 3 planes are then used to represent the 
final FP.

The uncertainties associated with the FP slope are also important.  
Feigelson \& Babu (1992) not only show that the standard formula 
(e.g.\ Bevington 1969) for calculating the uncertainties on the derived
slope and intercept is in general mathematically incorrect, but they 
also show that the derived asymptotic, analytic expressions 
for the variance will under-estimate the true 
variance when the sample size is small (approximately $\leq$50).
They therefore recommend the use of a bootstrapping procedure to 
gauge the variance on the fitted slope and its intercept. This approach 
was followed in this study.  However, this numerical resampling 
technique was modified for application to the small sample of Fornax 
elliptical galaxies in order to exclude 
10\% of the 400 bootstrapping simulations with a
negative, rather than positive, slope.  
These simulations were excluded from the computation 
of the final slope variance due to their obviously wild departure from the 
expected value.  Such FP's are due to the small sample size of 7 points 
in Fornax.  
With the somewhat larger sample of Virgo galaxies (25) this 
problem did not arise in any of the 2000 simulations used.  

The total sample of Virgo galaxies contains 18 E's and 7 S0's.  
It was therefore possible to compute the FP using both the entire
sample, and using only elliptical galaxies.  In addition, 
given the four sets of FP parameters 
($\log \sigma _{0}, \log R_{\rm e,4}, \Sigma _{\rm e,4}$),
($\log \sigma _{0}, \log R_{\rm e,n}, \Sigma _{\rm e,n}$),
($\log \sigma _{\rm tot,4}, \log R_{\rm e,4}, \Sigma _{\rm e,4}$) and 
($\log \sigma _{\rm tot,n}, \log R_{\rm e,n}, \Sigma _{\rm e,n}$), 
a total of eight FP's have been computed for the Virgo cluster.
The results are shown in Table~\ref{tab2} and Table~\ref{tab3},
where the rms residual is that perpendicular to the fitted plane. 
The uncertainties on the estimated slopes are those from the
bootstrapping procedure, and are therefore larger than the standard
analytically determined uncertainties derived for the FP.  
Using the sample of 7 available Fornax elliptical galaxies 
an additional four planes were computed using the above mentioned four
sets of parameters. Table~\ref{tab4} shows the results.

\begin{table}
\caption{Fundamental Plane $R$$\propto$$\sigma ^{A}\Sigma ^{B}$ for a sample of 25 E \& S0 Virgo galaxies constructed using the bisector method of linear regression in 3 dimensions.  The errors are from a bootstrapping procedure.  The approach referred to as $R^{1/4}[\sigma _{0}]$ has used the effective radius from the $R^{1/4}$ model, the mean surface-brightness within this radius, and the projected CVD, $\sigma _0$, in the construction of the FP.  $R^{1/n}[\sigma _{\rm tot,n}]$ uses the infinite aperture velocity dispersion term, $\sigma _{\rm tot,n}$, and the effective radius and associated surface brightness term from the $R^{1/n}$ model.}
\label{tab2}
\begin{tabular}{lccc}
\hline
model&\multicolumn{2}{c}{$R$$\propto$$\sigma ^{A}\Sigma ^{B}$}&$\perp$ rms \\[.2ex]
 &A&B&residual\\[10pt]
$R^{1/4}, \sigma _{0}$& 1.10$\pm$0.14 & -0.55$\pm$0.09 &0.099\\
$R^{1/n}, \sigma _{0}$& 1.12$\pm$0.17 & -0.64$\pm$0.07 &0.084\\
$R^{1/4}, \sigma _{\rm tot,4}$& 1.17$\pm$0.15 & -0.60$\pm$0.08 &0.100\\
$R^{1/n}, \sigma _{\rm tot,n}$& 1.37$\pm$0.16 & -0.76$\pm$0.05 &0.077\\
\hline
\end{tabular}
\end{table}

\begin{table}
\caption{Fundamental Plane $R$$\propto$$\sigma ^{A}\Sigma ^{B}$ for a sample of 18 E Virgo galaxies constructed using the bisector method of linear regression in 3 dimensions.  The errors are from a bootstrapping procedure.}
\label{tab3}
\begin{tabular}{lccc}
\hline
model&\multicolumn{2}{c}{$R$$\propto$$\sigma ^{A}\Sigma ^{B}$}&$\perp$ rms \\[.2ex]
 &A&B&residual\\[10pt]
$R^{1/4}, \sigma _{0}$& 1.19$\pm$0.21 & -0.60$\pm$0.11 &0.084\\
$R^{1/n}, \sigma _{0}$& 1.22$\pm$0.25 & -0.66$\pm$0.12 &0.071\\
$R^{1/4}, \sigma _{\rm tot,4}$& 1.28$\pm$0.24 & -0.65$\pm$0.11 &0.085\\
$R^{1/n}, \sigma _{\rm tot,n}$& 1.72$\pm$0.24 & -0.74$\pm$0.09 &0.050\\
\hline
\end{tabular}
\end{table}

\begin{table}
\caption{Fundamental Plane $R$$\propto$$\sigma ^{A}\Sigma ^{B}$ for a sample of 7 E Fornax galaxies constructed using the bisector method of linear regression in 3 dimensions.  The errors are from a bootstrapping procedure.}
\label{tab4}
\begin{tabular}{lccc}
\hline
model&\multicolumn{2}{c}{$R$$\propto$$\sigma ^{A}\Sigma ^{B}$}&$\perp$ rms \\[.2ex]
 &A&B&residual\\[10pt]
$R^{1/4}, \sigma _{0}$& 1.69$\pm$0.83 & -0.85$\pm$0.30 &0.064\\
$R^{1/n}, \sigma _{0}$& 1.79$\pm$0.79 & -0.98$\pm$0.28 &0.049\\
$R^{1/4}, \sigma _{\rm tot,4}$& 1.85$\pm$0.86 & -1.00$\pm$0.35 &0.070\\
$R^{1/n}, \sigma _{\rm tot,n}$& 2.03$\pm$0.78 & -1.07$\pm$0.30 &0.050\\
\hline
\end{tabular}
\end{table}

A Principal Component Analysis (PCA) was performed on each of the four 
3 dimensional data sets from the combined E \& S0 Virgo galaxy sample.
The code from Murtagh \& Heck \shortcite{MaH87} was implemented 
to show the degree to which the data sets are defined by a 2-dimensional
plane within the 3-space of observables.  Table~\ref{tab5} displays
the fractional variance of the data along the major eigenvectors
of the FP parameter space.  This reveals that 
greater than 97\% of the variance in the data indeed lies in a plane, 
confirming the appropriateness of constructing a plane to describe the 
properties of elliptical galaxies.

\begin{table}
\caption{Fractional Variance from the Principal Component Analysis on the sample of 25 Virgo E \& S0 galaxies.}
\label{tab5}
\begin{tabular}{lccc}
\hline
model&Var$_{1}$&Var$_{2}$&Var$_{3}$\\[10pt]
$R^{1/4}, \sigma _{0}$&76.03$\%$&21.99$\%$&1.98$\%$\\
$R^{1/n}, \sigma _{0}$&82.79$\%$&15.80$\%$&1.41$\%$\\
$R^{1/4}, \sigma _{\rm tot,4}$&72.30$\%$&25.57$\%$&2.13$\%$\\
$R^{1/n}, \sigma _{\rm tot,n}$&75.18$\%$&23.44$\%$&1.38$\%$\\
\hline
\end{tabular}
\end{table}

\section{The Virgo-Fornax distance-modulus}
\label{sectd}

Only the elliptical galaxy population is used in the following determination 
of the Virgo-Fornax distance-modulus due to the contaminating influence 
of the S0 galaxies on the FP.  Contaminating in the sense that S0 
galaxies may possess a significant degree of rotational energy which 
is ignored in the present analysis.

The problem of computing the offset between two parallel data sets, with
an emphasis on cosmic distance scale calibrations, 
is well described in Isobe et al.\ (1990) and Feigelson \& Babu (1992).  
If the objective in constructing the FP is for the purpose of computing distances
between galaxy clusters, then the FP parameters should be treated differently 
than they were treated in Section~\ref{sectc} (where the FP was constructed 
for comparison with a theoretical prediction).  
In this case, one wishes to predict the 
value of a  distance-dependent quantity, namely $\log R$, from the 
distance-independent quantities $\log \sigma $ and $\Sigma $.  
Consequently, the plane that is 
fitted to the data should be computed through an ordinary least-squares
minimisation of the distance-dependent quantity, rather than using the 
bisector method of linear regression which treats all variables equally
(Section~\ref{sectc}). 
This approach will minimise the uncertainty on the estimate of the 
relative distance between clusters.
Such an approach was taken with the two extreme Virgo and Fornax data sets,
($\log \sigma _{0}$,$\log R_{\rm e,4}$,$\Sigma _{\rm e,4}$) and 
($\log \sigma _{\rm tot,n}$,$\log R_{\rm e,n}$,$\Sigma _{\rm e,n}$).
(These are extreme in the 
sense of the `standard' versus `improved' treatment of galaxy structure 
and dynamics.)  
The best fitting planes, under this minimisation scheme, are given in
Table~\ref{tab6}, as are the rms residuals of the data projected along the 
$\log R$ axis. 

In order to use the methods in Feigelson \& Babu (1992), these 3-dimensional
data sets have been reduced to 2-dimensions by combining the 
distance-independent variables $\log \sigma $ and $\Sigma$.  These were
combined in such a way as to preserve the previously obtained optimal
linear regression solution.  This meant computing the `mixing' value
b=--$B/(2.5A)$, where $A$ and $B$ are the FP exponents given in 
Table~\ref{tab6}, giving the quantity $\log \sigma + b$$<$$\mu$$>$, 
where $<$$\mu$$>$ is the mean surface brightness in mag units.\footnote{$\Sigma$
is the mean surface brightness in linear units.}  
A further linear regression calculation was performed, minimising the
$\log R$ residuals at the expense of the ($\log \sigma + b$$<$$\mu$$>$) 
residuals.  The code SLOPES, discussed in Isobe et al.\ (1990), was used 
to do this, and the results are presented in Table~\ref{tab7}.  
As expected, due to the choice of the mixing value, the asymptotic formula 
gave the same slope as derived with the previous treatment of the FP variables 
in 3-dimensions.  The slopes obtained with bootstrap and jackknife sampling
are seen to vary slightly and have different associated uncertainties, as
expected for small sample sizes.  The errors from the
bootstrapping procedure for the Fornax data set have been restricted as 
discussed previously.

\begin{table}
\caption{Fundamental Plane $R$$\propto$$\sigma ^{A}\Sigma ^{B}$ using only elliptical galaxies.  Constructed using ordinary linear regression, in 3-dimensions,  that minimised the $\log R$ residuals.  The errors are from a bootstrapping procedure.}  
\label{tab6}
\begin{tabular}{lccc}
\hline
model&\multicolumn{2}{c}{$R$$\propto$$\sigma ^{A}\Sigma ^{B}$}&($\log R$) rms \\[.2ex]
 &A&B&residual\\[10pt]
Virgo: $R^{1/4}, \sigma _{0}$& 1.11$\pm$0.27 & -0.59$\pm$0.13 &0.134\\
Fornax: $R^{1/4}, \sigma _{0}$& 1.14$\pm$0.84 & -0.59$\pm$0.29 &0.120\\
Virgo: $R^{1/n}, \sigma _{\rm tot,n}$& 1.62$\pm$0.27 & -0.74$\pm$0.12 &0.100\\
Fornax: $R^{1/n}, \sigma _{\rm tot,n}$& 1.47$\pm$0.82 & -0.81$\pm$0.34 &0.110\\
\hline
\end{tabular}
\end{table}

\begin{table*}
\begin{minipage}{116mm}
\caption{Ordinary least-squares regression analysis for $\log R_{\rm e}$ versus ($\log \sigma +b\Sigma _{\rm e}$), using the program SLOPES (Feigelson \& Babu 1992).  In each case, the value of $b=-B/(2.5A)$ is derived from the best fitting plane shown in Table 6.}
\label{tab7}
\begin{tabular}{lccc}
\hline
method&Asymptotic Formula&Bootstrap&Jackknife\\[.2ex]
 &Slope (A)&Slope (A)&Slope (A)\\[10pt]
Virgo, $R^{1/4}[\sigma _{0}]$: b=0.21 & 1.11$\pm$0.08 & 1.10$\pm$0.08 & 1.11$\pm$0.09 \\
 & & & \\
Fornax,$R^{1/4}[\sigma _{0}]$: b=0.21 & 1.14$\pm$0.27 & 1.21$\pm$0.45 & 1.14$\pm$0.38 \\
 & & & \\
Virgo, $R^{1/n}[\sigma_{\rm tot,n}]$: b=0.18 & 1.62$\pm$0.09 & 1.62$\pm$0.09 & 1.62$\pm$0.10 \\
 & & & \\
Fornax, $R^{1/n}[\sigma_{\rm tot,n}]$: b=0.22 & 1.47$\pm$0.38 & 1.40$\pm$0.79 & 1.41$\pm$0.75 \\
\hline
\end{tabular}
\end{minipage}
\end{table*}

\begin{table}
\caption{Distance moduli given by 5($\Delta \log R$), where $\Delta \log R$
is the computed intercept offset between the cluster FPs, as described 
in the text.  These have been obtained using a generalisation of the 
Working-Hotelling confidence bands (Feigelson \& Babu 1992), using both 
Virgo and Fornax as the calibration sample.  
The uncertainties are from the asymptotic formula fits given in Table~\ref{tab7}.}
\label{tab8}
\begin{tabular}{lcc}
\hline
Calibrator&$R^{1/4}[\sigma _{0}]$&$R^{1/n}[\sigma_{\rm tot,n}]$\\[.2ex]
sample&mag.\ offset&mag.\ offset\\[10pt]
Virgo:& --0.46$\pm$0.23& --0.32$\pm$0.22\\
Fornax:& +0.45$\pm$0.16& +0.25$\pm$0.12\\
\hline
\end{tabular}
\end{table}

The two techniques described in Feigelson \& Babu (1992) for estimating
the offset of parallel data sets have both been implemented here.  
These methods are subject to the condition that the Virgo and Fornax 
elliptical galaxy population define FPs which are parallel to each other. 
Given the uncertainties on the derived FP slopes (Table~\ref{tab7}), 
the Fornax cluster data does indeed define a plane which is consistent 
with being parallel to the Virgo cluster data. 
This is true when $R^{1/4}$ model parameters and CVD's are used, and 
also when $R^{1/n}$ model parameters and infinite aperture velocity 
dispersions are used to construct the FP. 

In the following method of analysis, 
a linear regression is performed on one of the cluster data sets, and the 
resulting regression line from this {\it calibration} is then applied to the 
second (parallel) cluster. 
One technique for measuring the intercept offset between parallel data sets was 
pioneered by Working \& Hotelling (1929) and further developed by 
Feigelson \& Babu (1992) to permit the $x$-values (i.e.\ distance-independent
values) to be random variables rather than fixed quantities.  With this 
method, the calibrated regression line for Virgo was applied 
individually to the data points in Fornax and a weighted mean offset attained; 
the method is discussed
in Section 3.2 of Feigelson \& Babu (1992) and their Appendix A2.
This procedure encompasses the fact that the confidence intervals 
about the $y$-value one is trying to predict, given its $x$-value, will be  
larger for those data points at the ends of the distribution.  This is 
because  
the uncertainty in the slope of the calibration line/plane gives greater
freedom of movement to the calibration line at the ends of the distribution 
than in the center.  
This method of analysis gave an intercept offset, along the $\log R$ axis 
between the FP's defined by the Virgo and Fornax
data sets, of $\Delta \log R$=-0.092$\pm$0.045 dex and
-0.064$\pm$0.043 dex for the ($R^{1/4}$,$\sigma _0$) parameters and the
($R^{1/n}$,$\sigma _{tot,n}$) parameters, respectively.  

In a similar fashion, the Fornax cluster has been used as the calibration 
sample, where the fitted regression line for Fornax is applied to the Virgo 
cluster.  This was done with the Working-Hotelling confidence bands.  
Using the `standard' FP parameters 
($\sigma _{0}$,$R_{\rm e,4}$,$\Sigma _{\rm e,4}$), an intercept offset of 
+0.091$\pm$0.031 dex was computed.  Using the `improved' FP
parameters ($\sigma _{\rm tot,n}$,$R_{\rm e,n}$,$\Sigma _{\rm e,n}$), an 
intercept offset of +0.051$\pm$0.025 dex was computed.  
These intercept offsets translate into a distance modulus of +0.45$\pm$0.16 
mag (using $R^{1/4}$ model parameters and $\sigma _0$) and 
+0.25$\pm$0.12 mag (using $R^{1/n}$ model parameters and 
$\sigma _{tot,n}$).  The former estimate placing Fornax some (23$\pm$8)\%
further away than the Virgo cluster, compared to the latter estimate of
(12$\pm$6)\%. 
These solutions are fully consistent with those obtained using the Virgo
cluster as the calibration sample (see Table~\ref{tab8}). 

An alternative approach for computing the offset of parallel data sets
was achieved following the methods in Section 3.1 of 
Feigelson \& Babu (1992) and their Appendix A1.  
The FP data set from the Virgo cluster of galaxies was used to calibrate 
the {\it regression line} which was then applied to the FP data set of the Fornax 
cluster as a whole (and {\it vice-versa}).  This was done in order to derive 
the intercept offset, along the $\log R$ axis, between the FP's defined by
the Virgo and Fornax clusters. 
The solutions obtained were similar, within 0.001 dex, to those calculated
using the Working-Hotelling confidence bands, although the associated 
uncertainties were generally equal or larger. 
This served as a useful consistency check on 
the determination of the cluster-cluster distance modulus. 

With the Working-Hotelling confidence bands, 
the uncertainty on the intercept offset, and hence the distance estimate, 
is reduced by the square-root of the sample size one is applying the 
calibrated solution to.  
Consequently, applying the Fornax {\it regression} line to Virgo, with a sample
size of 18, results in a more accurate solution than achieved by applying 
the Virgo {\it regression} line to Fornax, with a sample size of only 7.
However, this effect does have to compete with the fact that the 
uncertainty on the regression line is larger for a smaller calibration
sample.  As the generalised Working-Hotelling confidence bands propagate 
the errors in the calibrated line, this effect cannot be avoided, and 
the size of these errors influence the associated 
errors for the intercept offset.  The lack of symmetry 
for the error estimates when using Virgo or Fornax as the 
calibrator is due to following:  The error estimates on the slope of the
regression line, which were propagated in the above analysis, were derived
from the analytical expression for these terms.  
However, the errors on the slope and intercept 
will be under-estimated by the analytical expressions when the number
of data points is small.  This is seen in the FP solutions shown in 
Table~\ref{tab7}, where the jackknife and bootstrapping methods of analysis 
provide a better estimate of the true size of these errors.  
Consequently, the error estimates on the previously given distance moduli 
should be multiplied by approximately the ratio of the bootstrapping
error (or jackknife error if this is larger) to the asymptotic error 
(Feigelson \& Babu 1992).  

In addition, it should be
noted that the FP slope uncertainties in Table~\ref{tab7} were estimated when 
the mixing value, $b$, was held fixed and assumed to have zero error.  
It should be remembered that it is the offset between two parallel planes, 
not two parallel lines, that is being measured.  In order to fully
estimate the uncertainties on the derived distance modulus, one must allow for
errors in the optimal FP slopes for all axes.  In so doing, one should return to 
the bootstrapping errors obtained with the 3-dimensional data set 
(Table~\ref{tab6}).  Working in
3-dimensions, the errors associated with the FP exponent $A$, are 
obtained by also letting the exponent $B$ (and hence $b$) vary, as 
permitted by the bootstrapping analysis of the data.
Therefore, the error obtained in the 3-dimensional analysis represents the 
true error estimate of the exponent $A$.  Allowing for this, the errors to 
the solution for the Virgo-Fornax distance modulus are again multiplied 
by the ratio of the 3-D bootstrapping errors to the 2-D asymptotic errors.  
This procedure does not effect the optimal solution, only the associated 
uncertainties.  The resulting solutions place Fornax (23$\pm$44)\% 
($R^{1/4}$,$\sigma _0$), and (12$\pm$36)\% ($R^{1/n}$,$\sigma _{\rm tot,n}$) 
further away than Virgo.

\section{Discussion}

In the interest of statistical completeness, galaxies with a CVD of less 
than 100 km s$^{-1}$ were not explicitly removed from the analysis.  Due to 
the greater scatter in the FP below this value, past studies have enforced 
this arbitrary cut-off in their galaxy sample.  However, it seems plausible to
attribute the higher level of scatter to an inadequate treatment
of the kinetic energy for these galaxies. 
As this is an issue which is addressed in this study, it made sense to leave
these galaxies in the analysis. 

Contributions from rotational energy to the total 
kinetic energy of a galaxy can in principle influence the slope of the
FP. Djorgovski \& Santiago (1993) noted that gains from 
including a rotational energy measurement would be offset by the 
introduction of additional measurement errors.  Indeed, a marginal
effect has been reported many times (Simien \& Prugniel 1992; Busarello, 
Longo \& Feoli 1992; Prugniel \& Simien 1994). An asymmetric trend
between the galaxy FP residuals and their maximum rotational velocity 
has also been found by D'Onofrio et al.\ (1996). 
More quantitative estimates from Prugniel \& Simien (1996a,b), using a 
sample of 400 galaxies, claim that 
rotation can account for 25$\pm$15\% of the departure of the FP from the 
virial expectation.   While this figure may be true for FP's constructed
using early-type galaxies (E and S0), which includes almost all FP's 
constructed to date, this figure can 
be reduced by the exclusion of S0 galaxies, known to have rotating disks.
This was hinted at in the study by Busarello et al.\ (1997), who used only 
{\it bona fide} ellipticals to obtain a contribution of 
15$\pm28$\% to the tilt to the FP due to galaxy rotational energies. 

In this paper, the FP was constructed using a culled set of galaxies, removing 
the S0s and keeping the Es.  Contributions from possible rotational 
support to the kinetic energy term are expected to be minimal in this 
study of the FP.  Four of the nine S0 galaxies which were excluded
did actually have CVD's less than 100 km s$^{-1}$, leaving only one 
galaxy in the remaining sample of 25 ellipticals with a CVD less
than 100 km s$^{-1}$.  It may indeed be 
that the increased scatter observed for the FP at the low velocity
dispersion end is due to the presence of S0 galaxies that 
harbour rotating disks with significant rotational energies that
are simply not measured through CVD measurements.  For the Virgo cluster, 
exclusion of these objects resulted in a reduction of some 15-35\% in the 
scatter about the FP, as evidenced by the rms residuals listed in 
Table~\ref{tab2} and Table~\ref{tab3}. 

For the Virgo cluster, exclusion of the S0's caused the final FP solution
(incorporating the observed range of galaxy profiles and implied dynamics) 
to change from 
$R_{\rm e,n}\propto \sigma _{\rm tot,n}^{1.37\pm0.16}\Sigma _{\rm e,n}^{-0.76\pm0.05}$
to
$R_{\rm e,n}\propto \sigma _{\rm tot,n}^{1.72\pm0.24}\Sigma _{\rm e,n}^{-0.74\pm0.09}$.
This suggests that the inclusion of the S0 population not only increases the
scatter about the FP (see Tables~\ref{tab2} and~\ref{tab3})
but is also partly responsible for the departure of
the FP from the relation $R\propto \sigma ^{2}\Sigma ^{-1}$, predicted by
the virial theorem.  This is again easily understood as due to the
neglect of rotational energies in the construction of the observed FP. 
It is therefore suggested that in addition to treating broken structural
and dynamical homology, the construction of the FP always adopt the
procedure of only using {\it bona fide} elliptical galaxies, unless one
allows for rotational energies.

It is found that by addressing 
the range of structural and dynamical profiles evident amongst the
elliptical galaxy population, the departure of the FP from the plane
predicted by the virial theorem is reduced, confirming the results in
Paper I.   In addition, the scatter of the data points 
about the FP is reduced by some 20\% for the sample of Fornax ellipticals
and by the same amount for the sample of Virgo E and S0 galaxies.  
For the sample of Virgo ellipticals, the mean rms scatter about the 
FP is reduced by 40\%  after treating the range of galaxy structures and
dynamics.   This is evident
in Tables 2, 3 and 4, which give the best fitting FP relation when one
does and does not treat broken structural homology, and when one does 
and does not consistently measure the kinetic energy of the galaxies
(i.e.\ uses $\sigma _{tot}$ or CVD's, respectively). 
The first row of these 3 tables gives the FP which was constructed assuming 
that all galaxies are described by an $R^{1/4}$ profile and uses CVD's 
to represent the kinetic energy of each galaxy; This procedure gives the  
greatest departures of the FP from the virial plane.
The final row in each Table gives the FP that was constructed allowing for 
a range of
structural profiles amongst the galaxies, as described by the $R^{1/n}$ 
model, and measures the kinetic energy of each galaxy in a consistent 
manner by using the computed velocity dispersion within an aperture of 
infinite radius; This procedure gives the closest agreement 
between the FP and the virial plane.  In fact, the formal solution for the 
FP of the Fornax cluster 
($R_{\rm e,n}\propto \sigma _{\rm tot,n}^{2.03\pm0.78}\Sigma _{\rm e,n}^{-1.07\pm0.30}$) 
is found to be in very close agreement with the 
prediction of the virial theorem, but does have large uncertainties 
associated with it because of the small number (7) of Fornax elliptical
galaxies in the sample. 
Overall, it is concluded that addressing the range of structural profiles
amongst the elliptical galaxies and better treating the galaxy dynamics
reduces both the departure of the FP from the virial plane and the scatter
about the FP.

In Paper I the Sersic light profile model  was fitted to a sample of 
26 E/S0 Virgo galaxies  using the aperture magnitude profile 
data from Bower et al.\ (1992).  
Of these, 18 galaxies are in common with the sample of 25 used here.  
Comparison between the galaxy model parameters obtained in each study is not 
straight-forward, however. 
The galaxy profile data used here 
is in the form of surface brightness measurements along the major-axis, 
where as Paper I uses circular aperture 
magnitudes.  Consequently, the models fitted in Paper I are
more heavily weighted by the inner profile, and by necessity 
incorporate the central parts of the galaxy light.  By contrast, 
the treatment of the surface brightness profiles here gives all data
points along the profile equal weight, and does not fit 
the central portion of the profile.  
In addition, the use of aperture magnitude data samples light  
from all position angles, and in so doing represents a mean light 
distribution for the galaxy as a whole.  The presence of ellipticity 
gradients will result in the major- and minor-axis
light profiles being described by different shape parameters so that
$n_{maj}$$\neq$$n_{min}$.  This is clearly evident in the 
range of shape parameters, $n$, for the major- and minor-axis, shown 
in the study of the Virgo ellipticals by Caon et al.\ (1993).  
Consequently, the mean light distribution obtained through the use of 
circular aperture magnitudes should not be expected to be identical 
to the light distribution along the major-axis. 
Finally, the data of Bower et al.\ (1992) was taken in 
the V-band, and that of Caon et al.\ (1993) in the B-band.  
Possible colour gradients, although expected to be small, may also 
contribute to differing results. 

\begin{figure}
\centerline{\psfig{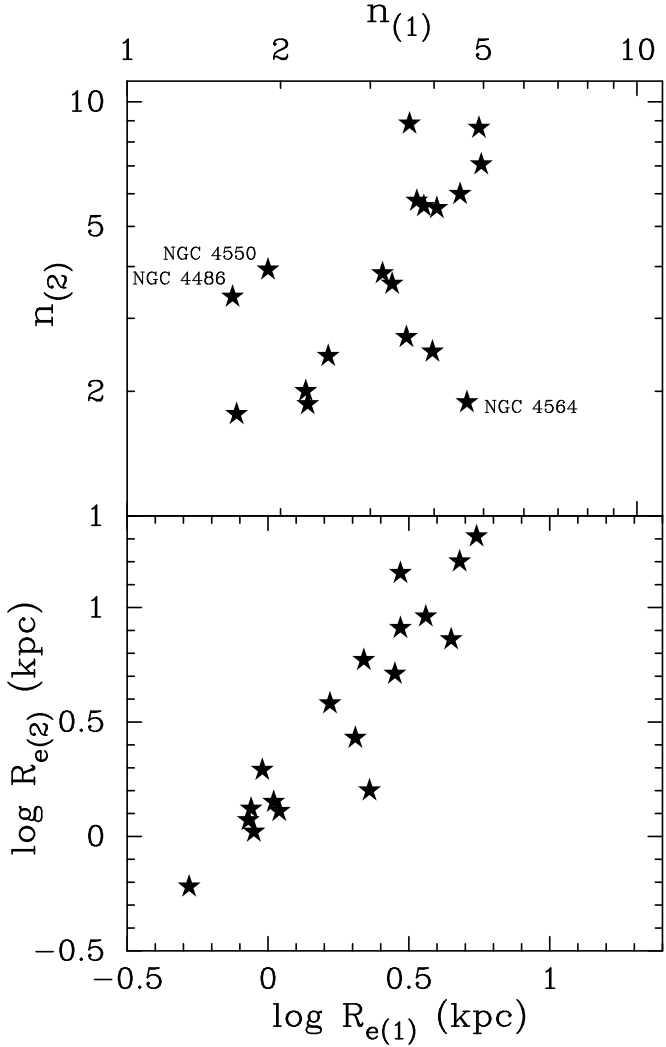}}
\caption{The Sersic $R^{1/n}$ model parameters for the effective 
radius, $R_e$, and shape parameter, $n$, are displayed for 18 galaxies
from Paper I which are in common with with the
sample in this study.  The subscript (1) denotes  that the 
parameters were obtained from Paper I, and the subscript
(2) refers to the values obtained in this study.}
\label{fig3}
\end{figure}

Despite the above factors, there is still fair agreement between the 
Sersic model parameters obtained using the different data sets. 
Galaxy scale-size and light profile shape 
for those galaxies in common between this Paper and Paper I 
are displayed in Figure~\ref{fig3}.  For comparison between the data sets, 
the Sersic model was fitted to the same radial extent ($<$40$\arcsec$) 
for each data set.  This outer radius cut-off comes from the aperture 
magnitude profile limit of Bower et al.\ (1992), and has been applied, 
for this comparison, to the major-axis profiles of Caon et al.\ (1994). 
The correlation seen between the shape parameters
has a linear correlation coefficient of 0.55, which increases to 0.82
upon removal of the 3 more discrepant points that are labeled in 
Figure~\ref{fig3}.  The correlation coefficient between the effective 
half-light radii is stronger at 0.93, suggesting that the range of galaxy 
sizes, and structures, can be {\it largely} reproduced from both types of 
observational data.  
The galaxies which appear to have the more discrepant shape parameters 
are the flattened S0 galaxy NGC 4550, NGC 4486, and NGC 4564 which has
a value of $n$=4.64 (Paper I) lying between $n_{maj}$=1.51
and $n_{min}$=5.08 (Caon et al.\ 1993) and is therefore not a discrepant
point after all.

Despite the above differences, it is interesting to compare the 
resulting FP's obtained using the independent data sets.  
Firstly, assuming structural homology and using $R^{1/4}$ model parameters 
and CVD measurements for the Virgo cluster of galaxies, the 
FP was found in Paper I to be described by 
$R_{\rm e,4}\propto \sigma _{0}^{1.33\pm0.10}\Sigma _{\rm e,4}^{-0.79\pm0.11}$.
This is to be compared with 
$R_{\rm e,4}\propto \sigma _{0}^{1.10\pm0.14}\Sigma _{\rm e,4}^{-0.55\pm0.09}$
found in this study. 
The exponents on the velocity dispersion term are consistent within the 
1$\sigma $ joint errors, and the exponents on the surface brightness
term are consistent within the 2$\sigma $ joint errors.
Allowing for broken structural homology and consistently 
measuring the velocity dispersion within an aperture of infinite radius, 
it was found in Paper I that the FP for Virgo is described by 
$R_{\rm e,n}\propto \sigma _{\rm tot,n}^{1.44\pm0.11}\Sigma _{\rm e,n}^{-0.93\pm0.08}$.
This should be compared with 
$R_{\rm e,n}\propto \sigma _{\rm tot,n}^{1.37\pm0.16}\Sigma _{\rm e,4}^{-0.76\pm0.05}$. 
The relative consistency between the exponents is the same as before.
It is noted that the uncertainties on the FP exponents are based upon the 
statistics of the linear regression to the data points; they 
do not incorporate possible measurement errors in the data points themselves.
It is also noted that the sample of Virgo galaxies used in Paper I 
contains 7 galaxies that this study does
not include, and this study also contains 7 galaxies not present in the sample
of Paper I. 
Using only those 18 galaxies in common, the resulting FP's were found to be in
acceptable agreement with each other, with the FP described by the data
in Paper I given by the relation
$R_{\rm e,n}\propto \sigma _{\rm tot,n}^{1.29\pm0.10}\Sigma _{\rm e,n}^{-0.89\pm0.05}$, 
and that obtained using this studies data given by the relation
$R_{\rm e,n}\propto \sigma _{\rm tot,n}^{1.26\pm0.12}\Sigma _{\rm e,n}^{-0.81\pm0.04}$.
The main point is that both studies show 
that by treating the range of luminosity and dynamical structures 
inherent amongst the early type galaxies the departure of the FP from
the virial plane is reduced. 

D'Onofrio et al.\ (1996) also constructed a FP for
the Virgo and Fornax clusters using the data of Caon et al.\ (1994). 
The approach taken in this study differs in several ways
to that of D'Onofrio et al.\ (1996). 
This work has used model-determined effective radii 
and the model-determined mean surface brightness within these radii, 
where as D'Onofrio et al.\ (1996) used model-independent values for these 
quantities (Caon et al.\ 1994).  The model-determined quantities 
are those derived from the $R^{1/n}$ model fitted to the undisturbed 
portion the galaxy light profile.  This portion is free from any 
peculiarities of the inner light profiles (noted by Caon et al.\ 1993), 
is not heavily distorted by the presence of a disk or dust lanes, 
and is uncontaminated from possible tidal interactions which may have
perturbed the original galaxy profile, particularly in the outer parts.  
The parameters used by D'Onofrio et al.\ (1996)
are better estimates of the half-light radii, in the true sense
of the term, in that they are the radii which enclose half of the
total light from the galaxy, although the total light was derived by extrapolating the 
observational data to $\mu _B$=32 with $R^{1/4}$ profiles (Caon et al.\ 1994), 
The preferred dynamical term used here differs from that used by 
D'Onofrio et al.\ (1996).  While they used CVD values, this paper
has also used, and in fact prefers,  the infinite aperture velocity dispersion
derived from the application of the Jeans equation (Ciotti 1991; Paper 1).
D'Onofrio et al.\ (1996) gave their Virgo FP solution as
$R_{\rm e}$$\propto$$\sigma _{0}^{1.26\pm0.09}\Sigma _{\rm e}^{-0.70\pm0.03}$.
This is in good agreement with the FP solution derived here, which
addressed the range of structural profiles but used CVD measurements.
For the sample of Virgo E and S0 galaxies the relation
$R_{\rm e,n}$$\propto$$\sigma _{0}^{1.12\pm0.17}\Sigma _{\rm e,n}^{-0.64\pm0.07}$
was obtained here.


D'Onofrio et al.\ (1996) computed a Virgo-Fornax distance modulus
of 0.50 mag (using the FP) and 0.45$\pm$0.15 mag (using the $D_n$-$\sigma $ 
relation).  
This result is in close agreement with the distance modulus derived
in this study based upon galaxy parameters from the $R^{1/4}$
profile and using CVD's, where a value of 0.45$\pm$0.16 mag (asymptotic error) 
was computed.  
Using $R^{1/n}$ model parameters and infinite aperture velocity dispersion
measurements, however, this study found the Virgo-Fornax distance modulus to be 
0.25$\pm$0.12 mag (asymptotic error).  
Given that D'Onofrio et al.\ (1996) allowed 
for different galaxy structures through using model-independent estimates
of the half-light radius and associated mean surface brightness, 
but used CVD's, the final estimate
of 0.25 mag appears to be different because of the treatment of the dynamical term
rather than the treatment of the structural quantities. 
Indeed, D'Onofrio et al.\ (1997)
introduced their own aperture correction to the velocity dispersion term, 
finding a reduced Virgo-Fornax distance modulus of 0.30$\pm$0.05 mag 
(formal error).

Comparison with other methods of distance determination yields good 
agreement with the distance modulus of +0.25$\pm$0.12 mag obtained in 
Section~\ref{sectd}.   The investigation by McMillan et al.\ (1993), 
using the planetary nebula luminosity function, found a Virgo-Fornax 
distance modulus of +0.24$\pm$0.10 mag.
%
%
Another accurate estimate has been derived using the method of 
surface brightness fluctuations, 
where a value of 0.20$\pm$0.08 mag has been obtained (Tonry et al.\ 1997).
Using a sample of E and S0 galaxies, Dressler et al.\ (1987) applied 
the $D_n$-$\sigma $
relation to compute a value of +0.14$\pm$0.18 mag.  An alternate 
approach using type Ia supernovae (Hamuy et al.\ 1991) found a yet 
lower value of 0.09$\pm$0.14 mag, although still consistent within the 
errors of the value derived here. 

However the result presented here is in poorer agreement with studies
using the globular cluster luminosity function (-0.5$\pm$0.2 mag, Bridges, 
Hanes \& Harris 1991) and with studies 
which have used the Tully Fisher (TF) relation for spiral galaxies. 
The H-band TF relation used by Aaronson et al.\ (1989) placed the 
Fornax cluster closer than Virgo, giving a distance modulus of 
-0.25$\pm$0.23 mag for Fornax relative to Virgo.  More recently, Bureau 
et al.\ (1996) used an I-band TF relation.  Demanding equal slopes
for the TF relation in each cluster, and allowing for errors on both axes,
they obtained a value of -0.04$\pm$0.15 mag. 
This result is not as inconsistent with the analysis in this paper as one 
might first think.  It should be noted that the bootstrapping error analysis, 
performed in Section~\ref{sectd}, revealed that the estimate of the error 
based upon analytical asymptotic expressions was too small due to the 
small sample size (see Table~\ref{tab7}).  Correction for this increased 
the error estimate from 0.12 mag to 0.27 mag.  The treatment in Bureau et al.\
(1996), and indeed most papers of this kind, has similarly under-estimated 
the true size of their errors for the same reason -- the analytical 
expression for the size of the errors under-estimates the true size of
the errors when the sample size is small (approximately $<$50).
In addition, it is known that spiral galaxies are 
less centrally concentrated within a cluster than the elliptical galaxies, 
so they may not provide as accurate a measurement
of the cluster distance as the ellipticals can.  This could be particularly
relevant for Virgo, a cluster known to have a considerable line-of-sight depth 
(Tonry, Ajhar \& Luppino 1990; Jacoby et al.\ 1992).

The Virgo-Fornax distance modulus obtained in this paper is also only 
marginally consistent with the study by
Bothun, Caldwell \& Schombert (1989) who used the surface brightness profiles 
of dwarf ellipticals to compute a distance modulus of -0.16$\pm$0.16 mag.
By the nature of their study, they also sampled a different galaxy 
population than used here, which again may partly explain the disagreement in
distance modulus.  
It is also observed that they made no mention of having used any numerical
resampling technique to estimate the size of their errors.   To
extrapolate, this opens
up the possibility that many past inconsistencies between cluster distance
estimations may be simply due to an inadequate, and under-estimation, of the
associated errors.

There are other effects which have been proposed to influence the tilt 
of the FP. One of these is orbital anisotropy in the galaxies' internal 
velocity fields.  This has been explored in detail in the
studies by Ciotti, Lanzoni \& Renzini (1996) and  Ciotti \& Lanzoni (1997).
Using the Jeans equation, they performed theoretical modeling of the dynamics
of a range of $R^{1/n}$ surface brightness profiles ($n$=1-10), and for 
the Hernquist (1990) and Jaffe (1983) density models. 
They explored possible variations in the orbital anisotropy with radius. 
Rejecting those models which are likely to be unstable, they found that
variations to the tilt of the FP due to orbital anisotropy are confined
within the known thickness of the FP and can thus have no significant 
influence on the FP tilt.

The influence of differing stellar populations, as given by line
strength indices such as Mg$_2$ or broad band colours such as B-V, 
was initially determined to have little or no effect on the slope of
the FP (Djorgovski \& Davis 1987; Lynden-Bell et al.\ 1988; J\o rgensen 
et al.\ 1993). 
However, more recent estimates of the influence from differing stellar 
metallicities and ages are suggesting a greater effect on the 
slope of the FP (Gregg 1992; Guzman \& Lucey 1993). 
Prugniel \& Simien (1996a,b) combined four broad band colours and the Mg$_2$ 
index in the extraction and computation of a mean measure of the influence 
of differing stellar populations.  They found that stellar populations 
may account for half of the observed tilt to the FP.
If confirmed, this would easily explain any remaining tilt to the FP, and
may place one in the unexpected position of having the FP tilt the other way.


\section{Summary}

For a sample of Virgo and Fornax early-type galaxies, the surface brightness 
profiles from Caon et al.\ (1994), which are known to represent a 
range of luminosity structures, have been used to explore the tilt and 
thickness of the FP.  This study also 
explores the range of luminosity structures present amongst the 
early-type galaxies using the Sersic $R^{1/n}$ model rather than
forcing structural homology through use of the $R^{1/4}$ model  
(see also Caon et al.\ 1993).
The $R^{1/n}$ light profile model has been used to parameterise
the size (as given by the effective radius $R_{\rm e,n}$) and mean
surface brightness ($\Sigma _{\rm e,n}$) for each galaxy.  
Solving the Jeans equation, the associated range of dynamical structures
has been computed and the projected velocity dispersion within an 
aperture of infinite radius (=1/3 of the virial velocity dispersion; 
Ciotti 1994) has been derived.  
This approach measures the velocity dispersion of each galaxy in a 
consistent fashion. CVD measurements sample different fractions
of different galaxy's dynamical profiles, where as the infinite aperture
velocity dispersion consistently measures galaxies in the same way, and
therefore provides a better estimate of a galaxy's kinetic energy.  

The trend of increasing shape parameter $n$ with increasing values of the
model-independent half-light radius (Caon et al.\ 1993), is also shown to be
present when using
the $R^{1/n}$ model-determined effective radius $R_{\rm e,n}$.
A $\chi ^{2}$ error analysis of the fitted model parameters shows that this
is not due to coupling between the model parameters implying
it is a real physical relation.
A principal component analysis reveals that greater than 97\% of the
variance in the FP data set ($\sigma $,$R$,$\Sigma $) resides within
2 dimensions, confirming the appropriateness of fitting a plane to this data.

It is concluded that broken structural and dynamical homology are responsible,
at least in part, for the departure of the FP from the relation
$R$$\propto$$\sigma ^{2}\Sigma ^{-1}$ predicted by the virial theorem.
Having used an independent set of data obtained by different observers with a
different instrument and reduction procedures, this confirms the results
of Paper I. 
This conclusion is supported by the additional analysis of the Fornax cluster.

For comparison of the the FP with the virial theorem, the FP was 
constructed using the bisector method of linear regression in 3 dimensions.
Due to the small galaxy sample size, a bootstrapping procedure was employed 
to estimate the uncertainty on the FP slope.  
Using a sample of 25 Virgo E (18) and S0 (7) galaxies, the FP constructed 
using $R^{1/4}$ model parameters and CVD measurements 
was found to be given by the relation 
$R_{\rm e,4}\propto \sigma _{0}^{1.10\pm0.14}\Sigma _{\rm e,4}^{-0.55\pm0.09}$.
Allowing for broken structural and broken dynamical homology, this plane
was observed to change to 
$R_{\rm e,n}\propto \sigma _{\rm tot,n}^{1.37\pm0.16}\Sigma _{\rm e,n}^{-0.76\pm0.05}$.
The perpendicular rms scatter about these FP's is 0.099 to 0.077 dex respectively. 

Given that the S0 galaxies likely possess significant rotational 
energy that is not dealt with through use of velocity dispersion
measurements, they were removed from the sample of Virgo galaxies
and the above analysis was re-performed.  
The resulting FP was found to change from
$R_{\rm e,4}\propto \sigma _{0}^{1.19\pm0.21}\Sigma _{\rm e,4}^{-0.60\pm0.11}$
to 
$R_{\rm e,n}\propto \sigma _{\rm tot,n}^{1.72\pm0.24}\Sigma _{\rm e,n}^{-0.74\pm0.09}$.  
The perpendicular rms scatter to the plane was found 
to decrease by 40\% from 0.084 to 0.050 with the fuller treatment of 
galaxy structure and dynamics. 

The FP for a sample of 7 elliptical galaxies in the Fornax cluster, was also 
computed under the same 
conditions as above.  The Fornax FP was found to change from
$R_{\rm e,4}\propto \sigma _{0}^{1.69\pm0.83}\Sigma _{\rm e,4}^{-0.85\pm0.30}$
to
$R_{\rm e,n}\propto \sigma _{\rm tot,n}^{2.03\pm0.78}\Sigma _{\rm e,n}^{-1.07\pm0.30}$, with perpendicular rms scatter of 0.064 and 0.050 respectively.
The formal solution of the latter FP is in close agreement with the 
plane predicted by the
virial theorem, and it will be of interest to see how/if this solution changes
for a larger sample of galaxies.  

Constructing the Virgo FP by minimising the residuals of the $\log R$ 
variable and applying this to the Fornax data, through use of the improved
Working-Hotelling confidence bands approach developed by Feigelson \& Babu
(1992), enabled an estimation of the relative distance between the two clusters.
Forcing structural homology (i.e.\ fitting $R^{1/4}$ profiles) and using
CVD's, a Virgo-Fornax distance modulus of 0.46$\pm$0.16 mag (analytical error) 
was obtained.  Allowing for the range of galaxy structures observed in the 
data, and using the infinite aperture velocity dispersion, the Virgo-Fornax 
distance modulus was computed to be 0.25$\pm$0.12 mag (analytical error).
A fuller estimate of the associated uncertainties to these distance moduli 
was achieved through a numerical resampling of the 3-dimensional FP data set. 
This analysis made the Fornax cluster (23$\pm$44)\% and (12$\pm$36)\% 
more distant than the Virgo cluster, for the respective models above.

\subsection*{Acknowledgments}

I thank Mauro D'Onofrio for making his galaxy profile data available 
to me in electronic form.
I also thank Eric Feigelson and Ron Kollgaard for providing me with their 
computer codes to compute the intercept offset between parallel data sets, 
and for their quick help with my queries.  
I am also grateful for the use of Eric Feigelson's computer code SLOPES.
I thank Matthew Colless for his proof reading of this manuscript 
and his useful suggestions.  
This research has made use of the NASA/IPAC Extragalactic Database (NED)   
which is operated by the Jet Propulsion Laboratory, California Institute   
of Technology, under contract with the National Aeronautics and Space      
Administration.

\end{document}